\documentstyle[12pt,aaspp4,epsfig]{article}

\received{19 November 1997}
\accepted{14 August 1998}
\slugcomment{To appear in ApJ (Letters); N.B.: unedited draft.}

\lefthead{Preece et al.}
\righthead{SSM Line of Death}

\begin{document}

\title{The Synchrotron Shock Model Confronts a \\
    ``Line of Death'' in the BATSE Gamma-Ray Burst Data}

\author{R.~D. Preece, M.~S. Briggs, R.~S. Mallozzi, G.~N. Pendleton, and 
W.~S. Paciesas}
\affil{Dept. of Physics, University of Alabama in Huntsville,
    Huntsville, AL 35899}

\and

\author{D.~L. Band}
\affil{Center for Astrophysics and Space Sciences, Code 0424, \\
    University of California at San Diego, La Jolla, CA 92093}

\begin{abstract}
The synchrotron shock model (SSM) for gamma-ray burst emission 
makes a testable prediction: that the observed low-energy power-law photon 
number spectral index cannot exceed $-2/3$ (where the photon model 
is defined with a positive index: $dN/dE  \propto E^{\alpha}$). 
We have collected time-resolved spectral fit parameters for over 100 
bright bursts observed by the Burst And Transient Source Experiment 
on board the {\it Compton Gamma Ray Observatory}. Using this database, 
we find 23 bursts in which the spectral index limit of the SSM is violated. 
We discuss elements of the 
analysis methodology that affect the robustness of this result, as 
well as some of the escape hatches left for the SSM by theory.
\end{abstract}


\section{Introduction}

Gamma-ray bursts (GRBs) have been an astronomical puzzle for 30 years. 
Details of the gamma-ray emission from bursts are 
still elusive, although there is an emerging agreement that persistent 
afterglow emission at other wavelengths from locations consistent 
with those of several recent GRBs can be related to expanding 
fireballs at cosmological distances (\cite{goodman86,metzger97,waxman97}). 
One promising mechanism that has been proposed for gamma-ray emission 
is the synchrotron shock model (SSM): synchrotron emission from particles 
accelerated in a relativistic shock is Lorentz-boosted into the gamma-ray 
band (\cite{rees92,meszaros93,reesmesz94,katz94,tavani96}). The SSM 
identifies the spectral break observed in most burst spectra with the 
characteristic synchrotron energy in the emitter's rest frame, boosted 
into the observer's frame. Thus, the fitted energy 
of the spectral break would contain information about both the bulk Lorentz 
motion of the emitters and the equipartition magnetic field producing the 
synchrotron radiation (\cite{tavani95}). In addition, the SSM makes the 
specific prediction that the low-energy power-law spectral index of the 
observed photon number spectra should not exceed $-2/3$, with the assumption 
that the optical depth of the shocked material is less than unity. We 
use the convention that the power-law index $\alpha$ takes a positive 
sign: $dN/dE \propto E^{\alpha}$. The $-2/3$ value 
is derived from the synchrotron single-particle emission spectrum: bulk 
relativistic motion ensures that the mean particle energy is large enough 
that there are few low-energy particles (\cite{katz94,tavani95}). Below the 
cyclotron fundamental energy, the low-energy spectrum is approximately that 
produced by mono-energetic particles and is thus independent of their actual 
distribution and should be a constant $-2/3$ spectral index power law. 
If that were the total story, the model would already be rejected, since it 
is well-known that bursts are observed to have a variety of spectral 
behaviors at low energies (\cite{band93,preece96,strohmayer97,crider97}). 
However, it also 
has been noted that the timescale for synchrotron cooling of the particles 
may be shorter than the duration of burst pulses (\cite{katz94b,sari96,sari97}). 
A cooling distribution of particles is characterized by a power-law 
index of $-2$, which translates into a $-3/2$ photon number index through the 
synchrotron power-law emission formula (\cite{rybicki}). 
If one includes the effects of cooling of the particle distribution, the 
low-energy spectral index can encompass the range of $-3/2$ to $-2/3$. 
However, the spectral slope still cannot be greater than the fundamental 
single-particle limit of $-2/3$.

In this {\it Letter}, we test the SSM limit on spectral behavior by examining 
how well the data support it. We draw our results from a catalog of 
time-sequences of spectral fits to 137 bursts selected for their 
high flux and fluence, using (mostly) Large-Area Detector (LAD) data from 
the Burst And Transient Source Experiment (BATSE) on the {\it Compton Gamma 
Ray Observatory} (see \cite{preece98} for details of the analysis). In the 
next section, we determine the best spectral indicators derivable from the 
data and show how they should bracket the `true' low-energy spectral 
behavior. In \S 3, for each of the bursts in the catalog we compare the effective   
low-energy power-law spectral index with the SSM limit line. Finally, in \S 4 
we discuss the implications these results have on further theoretical modeling.

\section{Spectral Modeling}

The model typically chosen for spectral fitting of bursts is the empirical 
`GRB' function (\cite{band93}):
\begin{eqnarray}
f(E) & = & A (E/100)^{\alpha} \exp{\biggl[\frac{-E(2+\alpha)}{E_{\rm peak}}\biggr]}\nonumber\\
{\rm if} \quad E & < & \frac{(\alpha-\beta)E_{\rm peak}}{(2+\alpha)} \equiv E_{\rm break} {\rm ,}\\
{\rm and} \quad f(E) & = & A \biggl[   %\{
\frac{(\alpha-\beta) E_{\rm peak}}{100(2+\alpha)}\biggr]^{(\alpha-\beta)}
\exp{(\beta-\alpha)} (E/100)^{\beta}\nonumber\\
{\rm if} \quad E & \geq & \frac{(\alpha-\beta)E_{\rm peak}}{(2+\alpha)}{\rm ,} \nonumber
\end{eqnarray}
where the four model parameters are: the amplitude A, a low-energy spectral index 
$\alpha$, a high-energy spectral index $\beta$ 
and an energy $E_{\rm peak}$ that corresponds to the peak of the spectrum in 
$\nu {\cal F}_{\nu}$ if $\beta$ is less than $-2$. In this expression, $E_{\rm 
peak}$ and $\alpha$ are jointly 
determined by the low-energy continuum, while $\beta$ is solely determined by 
the spectrum above the break energy, $E_{\rm break}$. 
If this break energy lies above the highest energy available to the detector, 
$\beta$ is ill-defined, so we must substitute a related form of the model 
with the high-energy power-law omitted. In cases where a sharp inherent 
curvature of the spectrum results in an unacceptable value of $\chi^2$ for a 
fit to the GRB spectral form, a broken power-law (BPL) model often generates 
better results.

\begin{figure}[t!]
\centerline{\epsfig{file=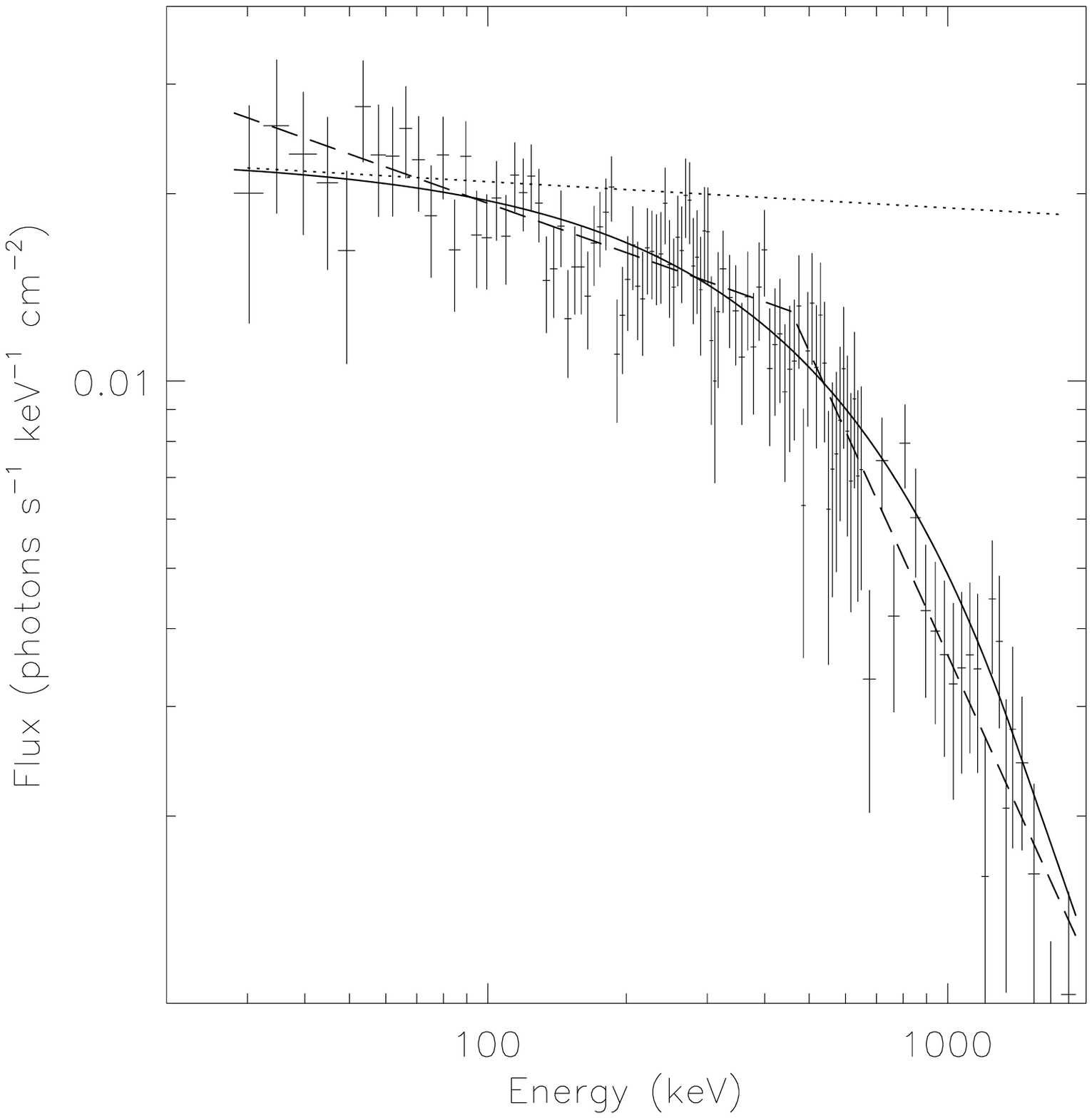,width=140mm}}
\caption[figure1.eps]{Photon spectrum accumulated between 0.448 -- 0.768 s 
after the trigger for 3B910814. Two best-fit models are plotted on top of the 
deconvolved data: the GRB function ({\it solid line}) and the BPL model 
({\it dashed line}; $\alpha_{\rm PL} 
= -0.263$). The tangent slope at 25 keV is also shown: $\alpha = -0.04$ 
({\it dotted line}). Neither model is as steep as $-2/3$ at low energies.
The errors represent model variances. 
\label{multi_purp}}
\end{figure}

Observed burst spectra are such that the data rarely approach the GRB spectral 
low-energy power-law within the energy range of BATSE and most other burst 
experiments. At the low end of the spectrum, $\alpha$ is approached only 
asymptotically, as seen by comparing the dotted line in Figure \ref{multi_purp} 
with the GRB model fit to an example spectrum ({\it solid line}). The model-dependent 
photon `data' rates consist of photon model rates weighted by the ratios of the 
deconvolved model rates and the count rates 
in each data channel, allowing different photon models to be represented simply 
on a single plot. A better measure of 
the actual lower-energy behavior is an effective spectral index, e.~g., the slope 
of the power-law tangent to the GRB function at some chosen energy ($E_{\rm fid}$), 
which can be found analytically:
\begin{equation}\
\frac{d \ln f(E)}{d \ln E} {\Bigg\vert}_{E=E_{\rm fid}} \equiv 
\alpha_{\rm eff}(E_{\rm fid}) = \alpha - 
(2 + \alpha) \frac{E_{\rm fid}}{E_{\rm peak}}{\rm .}
\end{equation}
For typical values of $\alpha$, i.~e., $\alpha > -2$, $\alpha_{\rm eff}(E_{\rm fid})$ 
will always be less than $\alpha$ by an amount that depends upon the value chosen for 
$E_{\rm fid}$. We chose $E_{\rm fid} = 25$ keV for the following reasons: The observed 
spectrum is consistent with the GRB function above about 25 keV; below this value, 
deviations from the standard GRB function have been observed that may indicate 
the existence of an additional low-energy spectral component (\cite{preece96}). 
In addition, 25 keV is just greater than the low-energy cut-off for the LADs; 
it is above an electronic spectral distortion in the SDs at typical gain 
settings (\cite{band92}) and it is also below fitted values for $E_{\rm peak}$ in 
nearly all bursts. The effective spectral index at 25 keV will be denoted 
$\alpha_{25}$. Once $E_{\rm fid}$ has been fixed, the correction gets smaller as 
the fitted value for $E_{\rm peak}$ gets larger, since $\alpha_{\rm eff}$ 
is closer to the asymptotic value $\alpha$. This is also seen in Figure 
\ref{multi_purp}, where $E_{\rm peak} = 1308$ keV, so $\alpha = -0.04 \approx 
\alpha_{25}$ (the GRB function is not appropriate for this spectrum and is shown 
for illustration only). 
In any practical case, $\alpha_{25}$ serves as an upper bound on the 
low-energy spectral index, in the sense that it is more positive 
than an average slope through the model. We will use $\alpha_{25}$ in 
this work, rather than $\alpha$, since it is more indicative of the 
slope actually seen in the data.

To the extent that all burst spectra have continuous curvature rather than a 
sharp break, a BPL function fit (Figure \ref{multi_purp} -- 
{\it dashed line}) will generate a spectral index (which we will call 
$\alpha_{\rm PL}$) that is a lower bound on the `true' low-energy spectral 
behavior, in that it will be more negative than it would be for a 
model that takes into account some curvature, such as the GRB function. Thus, 
to be conservative in our estimate for how well the BATSE data support the SSM, 
$\alpha_{\rm PL}$ provides the best available comparison, since if it lies above 
the limiting $-2/3$ line, we can be reasonably sure the `true' spectrum 
will violate the SSM. Of course, this will do us no good if the BPL model 
is not an acceptable fit to the data. Thus, we use fits to two models to put 
bounds on the `true' low-energy spectral behavior: $\alpha$ is a model parameter 
asymptotic to the data, but we can approximate the upper bound with the 
quantity $\alpha_{25}$ defined above. The BPL estimate, $\alpha_{\rm PL}$, 
is a lower bound.

\section{Observations}

In Figure \ref{max_alpha_lod}, we show results from our BATSE catalog of 
time-resolved spectroscopy of bright bursts (peak flux $> 10$ ph cm$^{-2}$ 
s$^{-1}$ and fluence $> 4 \times 10^{-5}$ erg cm$^{-2}$), using mostly LAD data. 
The value of $\alpha_{25}$ is plotted against $E_{\rm peak}$ for the spectrum in 
each burst for which $\alpha_{25}$ reaches its maximum value. In addition 
to the $\alpha_{25}$ -- $E_{\rm peak}$ parameters obtained by fitting the GRB 
function, we have also included $\alpha_{\rm PL}$ -- $E_{\rm break}$ pairs 
({\it diamonds}) for those bursts where use of the BPL model was more 
appropriate. Overlaying this plot is the 
synchrotron $-2/3$ spectral index `death line' ({\it dashes}), as well 
as the lower bound allowed for cooling distributions ({\it dotted 
line}). The SSM is quite safe within these boundaries; however, 
44\% of the total have maximum low-energy spectral indices above $-2/3$, 
in a region that is totally excluded in the SSM. In particular, 32\% of the 
BPL bursts have points that lie above this line. With some correlation 
between the two displayed parameters, the most stringent limits are set by the 
spectral indices of spectra with the highest values 
of $E_{\rm peak}$, as well as any bursts that have a BPL index greater than 
$-2/3$. Clearly, many such bursts exist; one is the example shown in Figure 
\ref{multi_purp}, which is a BPL fit with $\alpha_{\rm PL} = -0.263$ and 
$E_{\rm break} = 456$ keV, the highest-energy diamond above the $-2/3$ 
limiting line.

The error bars shown in Figure \ref{max_alpha_lod} represent the
1$\sigma$ error for each parameter considered singly, as obtained from
the covariance matrix of the fit. If the errors in the determination of
the spectral index were not normally distributed, perhaps having a broad
tail instead, then values that are much greater than the SSM limit of $-2/3$ 
might violate the SSM. To test this, we created 4 sets each of 1000 
simulated spectra, based on the GRB model spectrum propagated through 
the detector response, with random Poisson fluctuations determined for the 
counts in each data channel. These consisted of bright, medium and dim spectra, 
with expected errors of 0.06, 0.2 and 0.34, respectively, for the fitted value 
of $\alpha$. The parameter values assumed were E$_{\rm break} \sim 400$ keV, 
$\beta = -2$ and $\alpha = -2/3$, to provide a worst case test. A second set of 
medium-brightness spectra was also created, to test the effect of a small 
assumed value of E$_{\rm break} = 200$ keV. A histogram of the resulting fitted 
$\alpha$ values is approximately consistent with a Gaussian distribution for each set, 
however, distributions for the dimmer sets of spectra have slightly extended tails 
on the side of more positive values of $\alpha$. These larger values all have large 
computed errors associated with them, so the resulting distributions of deviations of 
$\alpha$ from the mean, weighted by their errors (see below), do not have a tail. 
Rather, the average over the 4 simulations for the number of error-weighted 
deviations greater than 2 $\sigma$ is 7.25, compared with the 22.7 deviations 
expected for 1000 trials in a normal distribution. 
Interestingly, the effect where large values of $\alpha$ have larger associated 
errors can be observed in Figure \ref{max_alpha_lod}, where the errors above the 
$-2/3$ line are, on average, 1.5 times those below. The KS probability we calculate 
from comparing the $\alpha > -2/3$ tails between our observations and the 
simulations is $3 \times 10^{-38}$ ($D_{\rm KS} = 0.26$ for $N_{\rm obs} = 1012$ 
and $N_{\rm sim} = 1689$), indicating that it is quite unlikely that the two 
distributions are the same.

\begin{figure}
\centerline{\epsfig{file=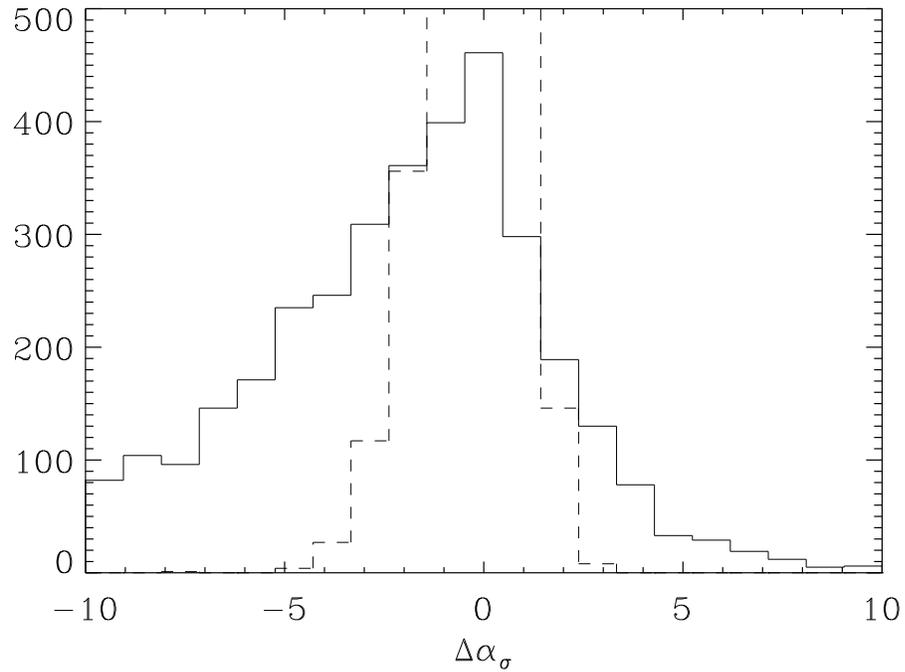,width=140mm}}
\figcaption[figure3.ps]{A histogram of deviations of the low-energy power-law 
indices from $-2/3$ in units of their $1\sigma$ error for 3957 fitted spectra. 
Positive deviations represent a violation of the SSM. Also plotted is the 
distribution of $1\sigma$ deviations from the mean for spectra indices obtained 
from fits to 4000 simulated spectra ({\it dotted}). The peak has been clipped so that 
the wings of the two distributions can be compared. \label{sig_dev}}
\end{figure}

Another way to view these results is as a histogram of 
$\Delta \alpha_{\sigma} \equiv (\alpha - (-2/3)) / \sigma_{\alpha}$ 
(here, we use $\alpha$ to stand in as a generic low-energy power-law index); that 
is, the deviation of $\alpha$ from $-2/3$ in units of the standard error for each fit. 
Figure \ref{sig_dev} shows the distribution of $\Delta \alpha_{\sigma}$ for the 
entire ensemble of 3957 spectra from the 137 bursts fit, where the SSM-violating 
region consists of the positive values to the right of zero. There are 312 spectra 
total in the bins containing $3 \sigma$ and greater (that is: $> 2.4 \sigma$). 
This may be compared with our expected error distribution, based upon the 4000 
total simulated spectral fits described above (Fig. \ref{sig_dev} -- {\it dashed 
line}), where the total over the same bins is 8 (in fact, $0 > 3.4 \sigma$). We may 
estimate the probability that 312 out of 3957 spectra would arise from 
$2.4 \sigma$ fluctuations larger than $-2/3$, assuming the SSM prediction 
that no such spectra exist. The simulations show that the chance probability of 
a $-2/3$ slope spectra having a fit value more than $+2.4 \sigma$ away from $-2/3$ 
is $\sim 8 / 4000$. Using the binomial probability distribution, the chance that 
the observed 312 out of 3957 could result from random fluctuations is 
$< 10^{-366}$, showing the observations to be quite inconsistent with the SSM. 

\section{Discussion}

\begin{figure}
\centerline{\epsfig{file=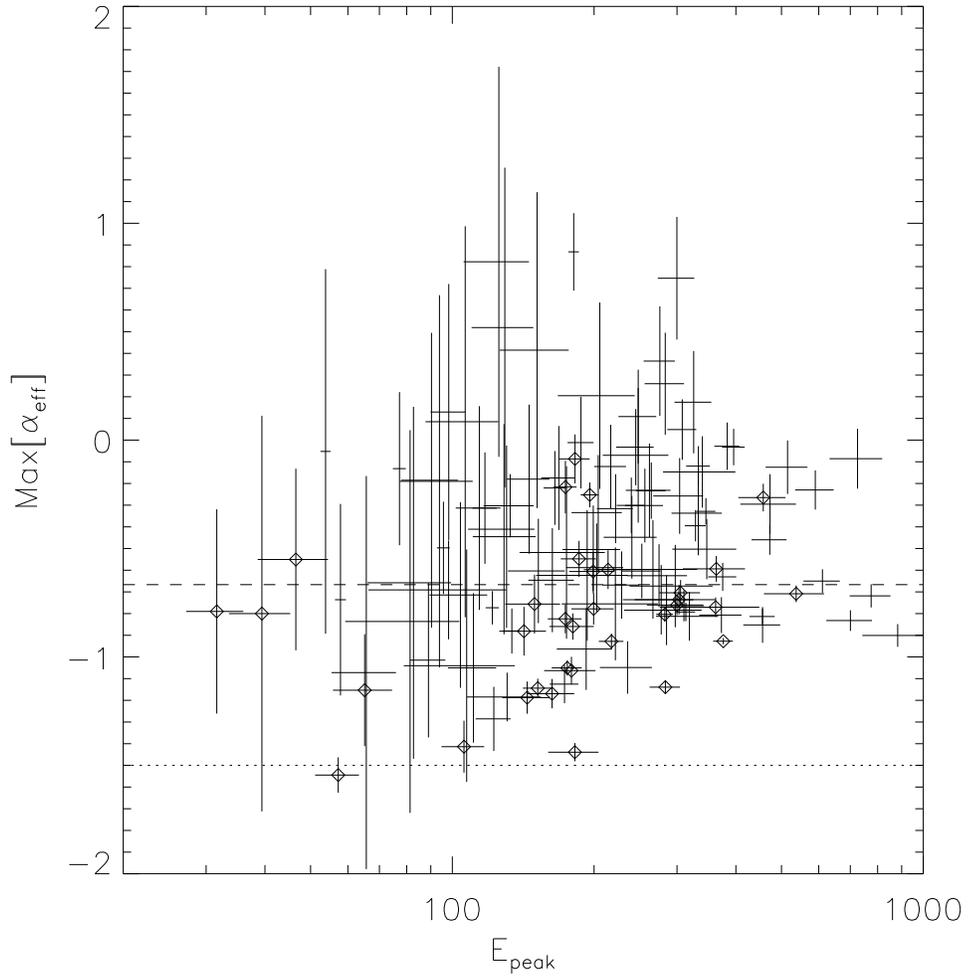,width=140mm}}
\caption[figure2.eps]{A plot of fitted low-energy power-law index against 
$E_{\rm peak}$. The spectrum chosen for each burst is the one with the largest 
value of $\alpha$. Diamonds indicate bursts fit with a broken power-law model. 
The SSM-violating region is bounded from below by 
the `death line' ({\it dashed}) and the acceptable region is bounded by the 
$-3/2$ line ({\it dotted}). Points where the error bars on $\alpha_{25}$ 
exceed the plot area were omitted for clarity. \label{max_alpha_lod}}
\end{figure}

We have shown that there are a large number of bursts that violate the 
limit on low-energy spectral behavior imposed by the basic synchrotron 
emission mechanism acting in a relativistic shock. It is worthwhile considering 
which spectral models can accommodate this observation. First of all, it has been 
suggested (\cite{liang97}) that Compton upscattering of soft photons by an energetic 
distribution of particles can significantly modify the basic synchrotron emission 
spectrum, with energetic particles boosting their own synchrotron emission into 
the observed gamma-ray band. The details of the spectral shape thus depend 
upon the particle distribution, as well as the shape of the photon spectrum 
at the energies where it is being sampled for upscattering. The observed steep 
low-energy spectral indices would arise for certain combinations of the source 
parameters and then would evolve to smaller values as the particles cool. An important 
prediction from this model is that a low-energy component, independent of the 
observed gamma-ray emission, must be present with sufficient strength to serve 
as a pool of photons for upscattering. It is indeed possible that this separate 
component may already have been observed (\cite{preece96,strohmayer97}). If the
low-energy portion is truly an independent component, it will have an independent 
time history as well. The troubling part of this idea is that {\it all} bursts 
should have this component to some extent, since there is no evident bi-modality 
to the low-energy behavior (cf. Figure \ref{max_alpha_lod}) that would indicate 
that the SSM-violating bursts are somehow different.

Synchrotron self-absorption (SSA) is another mechanism that would tend to increase 
the low-energy continuum spectral index. The maximum photon spectral index that 
could be observed is $+3/2$, so all of the bursts presented here would be 
consistent. The photon opacity must be close to unity in all these sources 
for some to be self-absorbed. For SSA to work the optical depth must be greater 
than one at energies below $E_{\rm peak}$. Thus, we trade one 
mystery for another: the narrow distribution of $E_{\rm peak}$ arises in the 
fact that many bursts have optical depths close to unity, rather than from a 
narrow distribution of Lorentz factors (see next paragraph). Also, if the photon 
density is high, it may be very difficult to overcome the 
opacity arising from photon-photon pair-production. 

Since the observed value of $E_{\rm peak}$ should scale as the bulk 
Lorentz factor of the emitting material to the fourth power, a narrow 
observed distribution of $E_{\rm peak}$ implies either that the rest-frame 
value is extraordinarily precisely determined for all bursts or that the Lorentz 
factors of the entire ensemble lie in a very narrow distribution. Since neither 
alternative is satisfactory, \cite{brainerd94} proposed that the observed 
spectra may arise from much different spectra that have been attenuated by 
intervening material through relativistic Compton scattering. The 
low-energy behavior is directly related to conditions at the source, namely 
the optical depth of the scatterers. Large optical depths will result in 
steep fitted low-energy power-law indices, and the distribution of these 
should somehow relate to the distribution of densities of the material 
surrounding putative sources.

\acknowledgments

Thanks to B. Schaefer who suggested adding error bars to the $\alpha$ -- 
$E_{\rm peak}$ figure from Preece et al. 1996. The anonymous referee has 
contributed suggestions that have lead to an improved error analysis.

\clearpage

\end{document}